\documentclass[preprint, prd]{revtex4}
\pdfoutput=1
\usepackage{bm}
\usepackage{amsmath,amssymb}
\usepackage[cal=boondox]{mathalfa}
\usepackage{graphicx}
\usepackage{float}
\usepackage{makecell}
\usepackage{siunitx}
\usepackage{cancel}
\DeclareMathAlphabet      {\mathbfit}{OML}{cmm}{b}{it}
\begin{document}

\begin{center}{\Large \textbf{
Measuring Unruh radiation from accelerated electrons
}}\end{center}

\begin{center}
G. Gregori$^{1*}$, G. Marocco$^{1,2}$, S. Sarkar$^{1}$, R. Bingham$^{3,4}$, C. Wang$^5$
\end{center}

\begin{center}

$^{1}$ Department of Physics, University of Oxford, Parks Road, Oxford OX1 3PU, UK
\\
$^{2}$ Lawrence Berkeley National Laboratory, 1 Cyclotron Rd, Berkeley CA 94720, USA
\\
$^{3}$ STFC Rutherford Appleton Laboratory, Chilton, Didcot, Oxon OX11 OQX, UK
\\
$^{4}$ Department of Physics, University of Strathclyde SUPA, Glasgow G4 0NG, UK
\\
$^{5}$ Department of Physics, University of Aberdeen, Aberdeen AB24 3UE, UK

(* Corresponding author: gianluca.gregori@physics.ox.ac.uk)
\end{center}

\begin{center}
\end{center}


\section*{Abstract}
{\bf
Detecting thermal Unruh radiation from accelerated electrons has presented a formidable challenge due not only to technical difficulties but also for lack of conceptual clarity about what is actually seen by a laboratory observer. We give a summary of the current interpretations along with a simpler heuristic description that draws on the analogy between the Unruh effect and radiation from a two-level atomic system. We propose an experiment to test whether there is emission of thermal photons from an accelerated electron. 
}


\section*{Introduction}

An era of accelerated expansion or `inflation' soon after the Big Bang is invoked to address several fundamental issues of the Friedmann-Lema\^{i}tre cosmology, viz. the horizon and flatness problems \cite{Baumann:2022mni}. Whereas these problems are not well-posed (as they require unjustifiable extrapolation back to $t=0$), it is widely accepted that the generation of a nearly scale-invariant spectrum of scalar density perturbations during inflation is crucial for seeding the growth of the observed large-scale structure in the Universe. The underlying physical mechanism is gravitationally driven particle production through quantum processes \cite{Parker:1968mv,Parker:1969qf,Parker:2012at}, which elegantly brings together general relativity and quantum field theory. The challenge of physically modelling such processes is however formidable, and here we consider whether it may be possible to \emph{experimentally} explore gravitational particle production using ultra-high intensity lasers. 




It is well known that particle-production phenomena can occur in a curved or dynamic spacetime \cite{Birrell:1982q,Fulling:1989nb,Parker:2009uva}. Famously, thermal radiation arises from particle production near the event horizon of a black hole, an effect known as Hawking radiation \cite{Hawking:1974rv,Hawking:1975vcx}. It had been noted earlier that particle production also occurs due to the varying gravitational field in an expanding Universe \cite{1939Phy.....6..899S,Parker:1968mv,Parker:1969qf,Parker:2012at}. (Although Schr\"odinger \cite{1939Phy.....6..899S} had remarked on this `alarming phenomenon' even earlier, his argument was not correct in detail \cite{Parker:1972vb}.) Of particular interest is the inflationary epoch in the early Universe when the Hubble expansion rate was very high and spacetime was very curved and dynamic. Understanding particle production during inflation 
should help to embed it in a physical framework \cite{Lyth:1998xn} and also elucidate non-thermal production of supermassive dark matter in the early universe \cite{Chung:1998zb}.

Simply put, an expanding space-time causes creation operators to evolve into superpositions of creation and annihilation operators. This so-called Bogoliubov transformation implies  the creation of particle-antiparticle pairs --- a consequence of the mappings between the Fock (particle number) states associated with different reference frames. This can happen in a curved spacetime or in an accelerating frame in flat spacetime \cite{Fulling:1972md,Davies:1974th}, both of which may be interpreted via the equivalence principle as due to gravitational particle production. 

Gravitational particle production has not been directly seen. Whereas primordial density perturbations are indeed observed via their imprint on the cosmic microwave background (CMB), to establish that these were generated during inflation requires detection of the concommitant primordial gravitational waves via their B-mode polarisation signature in the CMB; such a claim was made by the BICEP collaboration but subsequently shown to be unjustified \cite{Liu:2014mpa,BICEP2:2015nss}. However, soon to be commissioned multi-petawatt laser facilities \cite{Weber2017,gales2018} will enable us to access and control the unprecedented large acceleration they can impart to e.g. electrons which may lead to the observation for the first time of particle production in a non-inertial reference frame. This would constitute the experimental confirmation of the Fulling-Davies-Unruh effect \cite{Fulling:1972md,Davies:1974th,Unruh:1976db} which is considered to be equivalent to Hawking radiation as a consequence of black hole thermodynamics, and a pointer to quantum gravity. 

Our purpose here is to clarify what can be measured in such experiments and to guide  experimentalists with simple models that follow from basic principles. \color{black}
Any claim of detection of Unruh radiation needs to be an unambiguous measurement in the laboratory frame of an effect that tells us that thermal radiation is present in the accelerated frame. \color{black} 
We do not attempt rigorous derivations but rather aim to provide physically meaningful insights; 
this paper is an expanded version of our presentation at the Multi Petawatt Prioritization workshop \cite{mp3}. \color{black} We refer the reader to Crispino {\it et al.} \cite{Crispino2008} for an advanced discussion of Unruh radiation. \color{black} 

\section*{State of the art of our theoretical understanding}
As mentioned above, recent developments in ultra-high intensity lasers \cite{Strickland:1985c} have revived interest in the possibility of detecting both the Schwinger effect and testing non-perturbative QED effects
\color{black} \cite{Mourou:1998u,Bulanov:2003zz,Ahlers:2007qf,DiPiazza:2011tq,gonoskov.rmp.2022,Fedotov2023}. Projects such as the European Extreme Light Infrastructure \cite{Weber2017,gales2018} or the 
Shanghai Coherent Light Facility \cite{shen2018} \color{black}
will provide radiation beams with intensities exceeding $10^{23}\,\mathrm{\:W/cm^{2}}$. An electron placed at the focus of such beams should then experience an acceleration comparable to what it would feel at the event horizon of a $6\times10^{18}$ kg ($3\times10^{-12}\,M_{\rm \odot}$) black hole.
Indeed for such low mass black holes the surface gravity is strong enough that pairs of entangled photons can be produced from the vacuum, with one of the pair escaping to infinity. The black hole radiates and the spectrum has the blackbody form at the Hawking temperature \cite{Hawking:1974rv}. However astrophysical black holes are usually shrouded by the matter they are accreting. Given the  difficulties of directly observing Hawking radiation, Unruh proposed, by appealing to the Equivalence Principle (of gravitational and inertial accelerations), that a similar effect (hereafter called just the `Unruh effect') can be measured by an accelerated observer \cite{Unruh:1976db}.

While it is generally accepted that the semi-classical derivation of Hawking radiation is sound, it nevertheless makes use of several approximations that have not been tested.
Similarly, while an accelerated observer is believed to experience the equivalent Unruh radiation, it has been unclear what a detector in the laboratory frame will actually measure. Views are split, with some researchers believing that the answer is contained in  ordinary quantum field theory, while others point out that additional effects due to the acceleration must be included. A possible reason for such differing views is that defining what constitutes an experimental proof of Hawking and Unruh radiation requires a coherent understanding of rather disparate fields: viz. gravity, non-equilibrium quantum field theory in curved space-time, and high-intensity lasers.

    \begin{figure}
    \centering
    \includegraphics[width=0.5\linewidth]{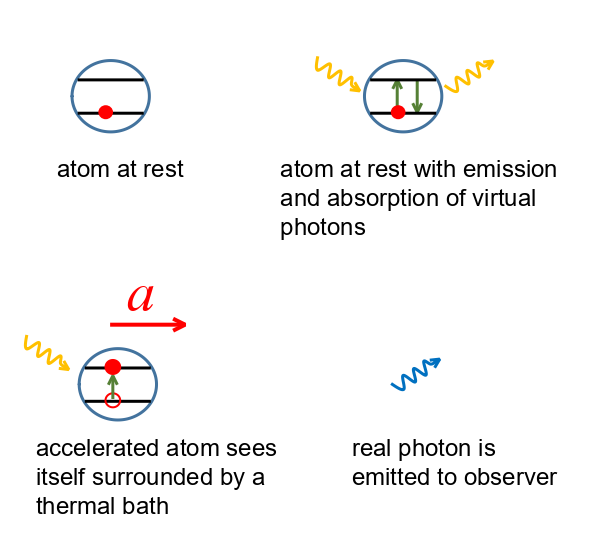}
    \caption{%
        Heuristic interpretation of the Unruh effect for a two-level atomic system. When the atom is at rest ({\it top row}), high order (virtual) processes involving the simultaneous emission and absorption of photons are permitted. However, when the atom is accelerated ({\it bottom row}), emission and absorption can become decorrelated leaving the atom in an excited state and emitting a photon to a distant observer.}
    \label{fig:Unruh}
    \end{figure}

The exact derivation of the Unruh effect is not simple, but  physical intuition can be gained by considering an idealized two-level atom (with energy separation $\Delta E = h \nu$) that is subject to a constant acceleration $a$ \cite{Scully:2017utk,Ben-Benjamin:2019opz}. In the accelerated state, higher-order processes by which a photon is simultaneously emitted and absorbed can also be considered (see Figure~\ref{fig:Unruh}). 
Since the acceleration incrementally changes the velocity of the atom, the frequency at which the photon is emitted can be slightly different than the frequency at which it is absorbed. 
However, if this change in frequency is larger than the linewidth of the atomic transition, the emitted photon disentangles from the absorbed one, and we are left with an atom in the excited state and a real photon being radiated away as shown in Figure~\ref{fig:Unruh}. By  Heisenberg's principle this occurs when $c^2/a \lesssim \ell$, where $\ell=c/\nu$ is the photon correlation length (i.e., wavelength), and we have assumed 
an acceleration time $t_\text{acc} \sim c/a$. This is the essence of the Unruh effect: 
it is a very general process which is associated with all accelerated bodies.

If the acceleration continues for a sufficiently long time (we will define what we mean by this later), the process of emission and absorption of photons by the atom reaches a steady state. The Unruh effect requires that the duration of the acceleration be at least comparable to the inverse of $A_{21}$, the Einstein coefficient for spontaneous emission (for the case of naturally broadened lineshape).
The frequency uncertainty due to the acceleration is
\begin{equation}
    \Delta \nu \gtrsim \frac{1}{2\pi t_{\rm acc}} = \frac{a}{2\pi c},
\end{equation}
so we can set $A_{21} \sim a/2\pi c$. 
\color{black}
Let us consider the accelerated frame, i.e., the instantaneous inertial frame in which the accelerated observer is at rest. In Rindler coordinates, frequencies ($\omega'$) and distances ($z'$) in the accelerated frame are related to those in the laboratory frame ($\omega$ and $z$, respectively) as: $\omega'(\tau) = \omega \, e^{-a \tau/c}$, $z'(\tau) = z \, e^{-a \tau/c} [1+\tanh{(a \tau/c)}]$, where $\tau$ is the proper time in the accelerated frame, \color{black} and the motion is in the $+$$z$ direction \color{black} \cite{alsing}. If we consider $\tau = 0$, then frequencies and distances in
the inertial laboratory frame and the accelerated frame are the same.
The spectral flux of emitted photons is given by
\begin{equation}
F^{\rm acc}_\nu({\tau=0}) \sim h \nu A_{21} \phi_\nu \ell^{-2} \sim \frac{h a \nu^3 \phi_\nu}{2 \pi c^3} ,
\end{equation}
where we have taken the photon correlation length to be the characteristic size of the emission volume, and $\phi_\nu$ is the line profile. 
\color{black}
In order to estimate the line profile, we notice that, because of the acceleration, frequencies that were emitted earlier in time are Doppler shifted. The Lorentz boost that relates the frequencies in the accelerated ($\nu'$) and inertial ($\nu$) frames is
\begin{equation}
    \nu'(\tau) = \nu \, \frac{1-v/c}{\sqrt{1-v^2/c^2}},
\end{equation}
where the accelerated frame moves with velocity $v(\tau) = c \tanh{(a \tau/c)}$ on the Rindler's trajectory.
Thus, the probability to emit a photon in the frequency range between $\nu'$ and $\nu'+d\nu'$ is simply related to the fractional change in velocity; hence $\phi_\nu \, d\nu' = dv/c$. This implies,
\begin{equation}
    \phi_\nu \, d\nu' = \frac{d v}{c \, d\nu'} \, d\nu' \sim \frac{1}{\nu} \, d\nu',
\end{equation}
where the last relation is obtained by assuming that most of the emission occurs near $\tau=0$.
We can thus approximate \color{black} $\phi_\nu = 1/\nu$, giving
\begin{equation}
F^{\rm acc}_\nu \sim \frac{h a \nu^2}{2 \pi c^3}.
\label{Fnu}
\end{equation}
\color{black} We notice that this relation should only be applicable for small frequencies since it relies on the assumption that most of the spectrum is built around $\tau=0$. \color{black}

As discussed in Ref. \cite{alsing}, the overall frequency spectrum that results from these Doppler shifts is thermal.
In the final step of this heuristic derivation of the Unruh effect, let us assume that the flux of emitted photons can be represented in terms of a blackbody function at some temperature $T_\text{U}$. In the low-energy regime, this is simply given by the Raleigh-Jeans formula:
\color{black}
\begin{equation}
F^{\rm acc}_\nu = \frac{2 \pi \nu^2}{c^2} k_\mathrm{B} T_\text{U}.
\label{FnuRJ}
\end{equation}
\color{black}
Equating the last two expressions, we get
\begin{equation}
T_\text{U} = \frac{\hbar a}{2 \pi k_\mathrm{B} c},
\end{equation}
which is just the Unruh temperature \cite{Unruh:1976db}
\color{black} (please note that in the definition of $T_\text{U}$ we have introduced the reduced Planck constant, $\hbar$, instead of $h$). \color{black}

From the considerations above we see that the key element in the realization of the Unruh effect is the atom having discrete energy levels, i.e. a system that can change its state or its internal energy. We call such a system a {\it detector}. Any particle with internal structure, e.g. a proton or a neutron, is a detector according to this definition. If we describe the two level detector as a classical harmonic oscillator, the damping rate is obtained by comparing the correlation length of the system ($\lambda$) to the \color{black} spatial \color{black} scale of the transition ($c/\nu)$, i.e.,
$\gamma_{cl}=(32 \pi^3/3) \, \nu^3 (\lambda/c)^2 \sim \nu^3 (\ell/c)^2$ \cite{Fitzpatrick,Ginzburg1987}, where $\lambda \sim \ell \sim c/\nu$ for such a transition.
Using this, the coefficient of spontaneous emission is \cite{Bhattacharya2012,Ben-Benjamin:2019opz}:
\begin{equation}
A_{21} \sim \frac{g^2}{4\pi} \gamma_{cl} \sim
\frac{g^2}{4\pi} \left(\frac{a}{2\pi c} \right),
\label{eqA21}
\end{equation}
where, for generality, we have introduced a coupling constant, $g^2/4\pi$. For discrete transitions of a detector involving the emission and absorption of photons and a change of internal state, $g^2/4\pi = 1$. If the transition is associated with an electron jumping to a different atomic level, this is described by the electromagnetic coupling $g^2/4\pi = \alpha$, where $\alpha$ is the fine structure constant.
The arguments here can be easily generalised. In Figure~\ref{fig:Unruh} we see that the accelerated atom can, in principle, couple to the emission and absorption of other particles e.g. gravitons and neutrinos \cite{matsas1999,Vanzella2001,Ahluwalia2016,Blasone2020}, and possibly even those beyond the Standard Model such as axions, dark photons or millicharged particles \cite{Raffelt:1987im,Wadud:2016pwu}.

\section*{Larmor power and radiation reaction}
A direct observation of the Unruh effect as outlined above would require the acceleration of atoms. However, because of their larger mass compared to that of electrons, the acceleration of atoms or ions is challenging. Using radiation pressure acceleration \cite{Gelfer.2016}, even at intensities $I_\text{L} \sim 10^{25}\,\mathrm{\:W/cm^{2}}$, we expect only $T_\text{U} \sim 10^{-3}$ eV. This temperature would indeed be high enough to affect the ionization state of the bound levels in the accelerated atom, hence the measurement of the ionization balance or line emission spectral changes from the atom as function of the acceleration 
(particularly for rotational/vibrational spectra of molecules) would provide possible ways for the detection of Unruh radiation. However the required laser intensity is well above what is achievable at any current laser facility. In addition, at $I_\text{L} \sim 10^{25}\,\mathrm{\:W/cm^{2}}$ QED effects such as copious production of electron-positron pairs \cite{Bell:2008zzb} would already make any detection of changes in the ionization balance very challenging. As discussed in Ref.~\cite{Stargen:2021vtg}, for small accelerations, the Unruh signal from an atom inside a cavity can resonate with its normal modes, enhancing the overall emission. This may be a promising avenue for a future experiment; however, the experimental realization of such a detector is still far from being realised. 

For all of these reasons, Unruh and others have instead adopted an alternative
approach to exploit the mathematical analogy between trans-sonic flowing water \cite{Weinfurtner:2010nu}, Bose-Einstein condensates \cite{MunozdeNova:2018fxv} and other analogue systems \cite{Drori:2018ivu} and the behaviour of quantum fields in the vicinity of a black hole horizon. While these experiments have successfully demonstrated the mathematical soundness of Hawking's solution, they may have fallen short in proving that the radiation is actually emitted by non-inertial bodies and
that the underlying theory is indeed physically correct \cite{Crowther}.

Electrons, instead, are much easier to accelerate so are the ideal `detector' to test the Unruh predictions. Large accelerations can already be realised in the laboratory using  high intensity lasers, e.g. lasers with intensity $I_\text{L} \sim 10^{19}$ W/cm$^2$ can accelerate electrons to an Unruh temperature of about 1 eV \cite{Chen:1998kp}. While, in principle, this can be measured in the laboratory, there is a fundamental issue that has seldom been discussed in the literature: the electron is a \emph{point} particle with no internal structure. This implies that what we mean by Unruh radiation for an accelerated electrons must be carefully defined,
\color{black} see, in particular Refs. \cite{Bell.1983o45,Bell1987,habs2006}. 
\color{black}
Bell and Leinaas \cite{Bell.1983o45,Bell1987} have pointed out that by treating the electron spin as a two-level quantum system interacting with the Unruh bath, the effect of the radiation field along the accelerated orbit of the electrons is different from the effect on an electron sitting at rest. 
This leads to a small, but non-vanishing depolarization of the electron beam. These conclusions have, however, been questioned \cite{jackson1,jackson2}. Moreover, the Unruh temperature in arbitrary trajectories \cite{gine,milgrom} may not be the same as for uniform linear acceleration, as it is commonly calculated.
Recently the possibility of accelerating an electron in a magnetic field has been considered \cite{Wang:2022ins}. While such a system exhibits finite transitions set by the Landau levels and thus its behaviour resembles that of an atom, its realisation would still push the limits of the currently available laser and magnetic field technology. 

Already the above discussion shows how controversial the subject of Unruh radiation has been in the literature \color{black} \cite{Narozhny.2002,Ford:2005sh, Cruz2016} \color{black}, especially concerning how to distinguish it from other classical and quantum radiation processes involving acceleration of charged particles.

Here we address these issues in a unified approach, with the aim of clarifying different interpretations of what is meant by radiation from an accelerated electron. While the electron has no internal energy states, its interaction with the accelerated vacuum (the Unruh photons) results in continuous changes of its momentum via electromagnetic scattering \cite{Landulfo:2019tqj,Hegelich:2022zca}.
First, for electromagnetic scattering, we take $g^2/4\pi = \alpha$ in Eq.~(\ref{eqA21}).
These energy changes can be infinitely small, so $\nu \rightarrow 0$ and 
$A_{21} \rightarrow 0$. This means that those small continuous transitions require infinitely long times to equilibrate and consequently the power radiated by the Unruh effect at those frequencies is very small. Eq.~(\ref{eqA21}) still remains valid if $\gamma_{cl} \lesssim a/2 \pi c$, implying that the uniform acceleration must continue at least for a time
\begin{equation}
t \gtrsim t_\mathrm{eq} = \left( \frac{64 \pi^4 h^2}{3 m^2 c^3 a} \right)^{1/3},
\label{teq}
\end{equation}
where $t_\mathrm{eq}  \sim 1/\nu$ is the thermalization time and we have taken $\ell = \lambda_C$, where $\lambda_C=h/m c$ is the Compton wavelength, i.e. the correlation length of the structure-less detector.
\color{black} Also, if we want to restrict our analysis to only non-relativistic motion, we should also require $t \ll t_{\rm acc}$, implying
$t \ll c/a$. \color{black}


We conclude this section by noting that the power emitted in Unruh radiation \color{black} in the accelerated frame is 
\begin{equation}
P^{\rm acc}_\nu = \frac{F^{\rm acc}_\nu \ell^2}{t_{\rm acc}} \sim {\cal B} \left(\frac{\hbar}{c^2}\right) \frac{g^2 a^2}{4\pi},
\end{equation}
\color{black}
where the factor ${\cal B} \sim 2/3$ arises from the fact that transitions where the detector undergoes an energy gain or an energy loss are both possible (hence a factor of 2), and because of the isotropy of space in three dimensions (giving a factor 1/3). \color{black} While $P^{\rm acc}_\nu$ is the power emitted in the accelerated frame, the same result would have been obtained had the calculation been done in the laboratory frame. Hence $P_\nu^{\rm lab} = P_\nu^{\rm acc}$,
\color{black} which is a somehow trivial argument for inertial frames since it is well-known from classical electromagnetism that the Larmor formula for the radiated power carries no net momentum and therefore is Lorentz invariant. On the other hand, one could also argue that the same considerations should apply for the frames that are instantaneously at rest with the accelerated charge.
Since Eq.~(\ref{Fnu}) was derived for slow-moving observers, the above relations are strictly valid only for non-relativistic motion.

However, the relation $P_\nu^{\rm lab} = P_\nu^{\rm acc}$ can be proven exactly using a full relativistic quantum field theory calculation of the emitted power by the accelerated charge which is performed in both frames \cite{vacalis2024}. It is shown \color{black} that the Unruh effect involving an accelerated electron reduces, at tree level, to nothing other than the classical Larmor radiation as seen in the laboratory frame. For the case of an electron, assuming that it had sufficient time to thermalize, we indeed reproduce the classical Larmor formula for electromagnetic radiation, which can be seen as the limit power achieved by accelerated detectors with no internal structure.

This appears to be a quite general result. 
Refs.~\cite{higuchi1992,higuchi1992b,Paithankar2020,Lynch:2019hmk} show that the rate of electromagnetic bremsstrahlung emitted by a static charge in the accelerated frame is the same as that measured by the inertial one. A similar conclusion for scalar fields was also reached by Refs.~\cite{Ren1994,Diaz2002,Cozzella:2020gci}. 
All the above authors conclude that the classical Larmor radiation of an accelerated charge corresponds to the emission and absorption of zero-energy Rindler photons of the thermal bath in the accelerated frame. The two processes are in fact exactly equivalent, the only difference being that in the Unruh case the calculation is performed in the accelerated frame, and for Larmor radiation in the laboratory frame.
\color{black}

\color{black}The equivalence between Larmor and Unruh radiation only \color{black} stands if we consider the lower order (classical) limit implied by the spectral flux in Eq.~(\ref{Fnu}). Going beyond the Rayleigh-Jean regime, Eq.~(\ref{FnuRJ}) becomes
\color{black}
\begin{equation}
F_\nu^{\rm acc} = \frac{2 \pi h \nu^3}{c^2} 
\frac{1}{e^{h \nu / k_\mathrm{B} T_\text{U}}-1},
\end{equation}
\color{black}
which then allows us to include the full energy spectrum of the emitted photons by a non-relativistic electron.
\color{black}
The above expression assumes that the detector experiences no recoil. However, as photons rescatter with the detector, the latter does experience a finite recoil $\mu$. The occupation of states described by the Planck function must then be changed to \cite{Wurfel2000,Lynch:2019hmk}
\color{black}
\begin{equation}
F_\nu^{\rm acc} = \frac{2 \pi h \nu^3}{c^2} 
\frac{1}{e^{(h \nu + \mu) / k_\mathrm{B} T_\text{U}}-1},
\end{equation}
\color{black}
where $\mu=\hbar^2 k^2/2 m$ plays the role of a chemical potential. Here $m$ is the mass of the detector and $\hbar k$ represents the recoil momentum. Since the average photon energy is of order of $k_\mathrm{B} T_\text{U}$, then $\hbar k c \sim k_\mathrm{B} T_\text{U}$, and we can approximate
$\mu \sim (k_\mathrm{B} T_\text{U})^2/2 m c^2$.
\color{black} To limit to sub-relativistic energies, we also require that the acceleration time is long compared to the time it takes a photon to cross a Compton wavelength, i.e., 
$t_\mathrm{acc} \gg \lambda_C/c$, which implies $k_\mathrm{B} T_\text{U} \ll m c^2$.
\color{black}
Regarding $\mu$ as a small correction, we have
\color{black}
\begin{equation}
F_\nu^{\rm acc} = \frac{2 \pi \nu^2}{c^2} k_\mathrm{B} T_\text{U} 
\left[1 - \frac{(k_\mathrm{B} T_\text{U})^2}{2 m c^2 h \nu} \right] \sim  \frac{h a \nu^2}{2\pi c^3} 
\left(1 - \frac{k_\mathrm{B} T_\text{U}}{2 m c^2} \right),
\end{equation}
\color{black}
where in the last step inside the parenthesis we have taken $h \nu \sim k_\mathrm{B} T_\mathrm{U}$. Using the same reasoning as above, the power emitted by Unruh radiation is
\color{black}
\begin{equation}
P_\nu^{\rm acc} \equiv P_\nu^{\rm lab} \sim {\cal B} \left(\frac{\hbar}{c^2}\right) \frac{g^2 a^2}{4\pi}\left(1 - \eta \frac{k_\mathrm{B} T_\text{U}}{2 m c^2} \right),
\label{qrr}
\end{equation}
\color{black}
where again ${\cal B} \sim 2/3$ and $\eta$ is a coefficient that depends on the details of the detector. For electrons, $\eta=24$ \cite{Lynch:2019hmk}. 
The above formula shows that recoil (quantum) corrections to the Larmor formula contain terms of order $a^2 k_\mathrm{B} T_\text{U}/m c^2$ \cite{Lin:2005uk,Lynch:2019hmk}. 
This is what is usually referred to as the Unruh effect for accelerated electrons \color{black} \cite{habs2006,oshita2016} \color{black}.
In addition, while relativistic corrections to the Larmor formula are outside the scope of the present work, we should stress that those too may further modify Eq.~\ref{qrr} in a non trivial manner. 
\color{black}

If we now use Eq.(\ref{qrr}) and repeat the same calculation used to derive the Abraham–Lorentz formula for the radiation reaction force \color{black} in the laboratory frame, we obtain:
\begin{equation}
{\bf F}_{RR}^{\rm lab} \sim {\cal B} \left(\frac{\hbar}{c^2}\right) \frac{g^2}{4\pi}\left(1 - \eta \frac{k_\mathrm{B} T_\text{U}}{m c^2} \right) \mathbfit{\dot a}.
\label{rr}
\end{equation}
\color{black}
This provides a first-order quantum correction to the non-relativistic Abraham–Lorentz radiation reaction force.
Recently, it has been argued \cite{Lynch:2019hmk} that the observation of quantum corrections to radiation reaction are an experimental proof of the Unruh effect. Indeed, quantum radiation reaction effects have been seen in aligned crystals \cite{Wistisen2018} as well as with high power lasers \cite{Cole:2017zca}.
However, since quantum corrections to the Abraham–Lorentz formula (and its relativistic counterpart) can also be derived directly from QED \cite{DiPiazza:2011tq} without appealing to the Unruh effect, the claim \cite{Lynch:2019hmk} 
\color{black} alone is not sufficient
without a firm theoretical understanding of the relationship between radiation reaction in the laboratory frame and higher-order corrections to the Unruh effect in the accelerated frame. \color{black} Unfortunately, the noise in the experimental setup was too high to allow the outcome to be established unambiguously.  
\color{black}

\section*{Avenues to the detection of the Unruh effect}
\color{black}
To proceed further we need to turn back to the radiation reaction formula
of Eq.(\ref{rr}). 
The quantum recoil correction implies a modified fine structure constant \cite{Lynch:2019hmk}
\begin{equation}
\alpha \longrightarrow \alpha \left(1 - \eta \frac{k_\mathrm{B} T_\text{U}}{m c^2} \right).
 \label{alpha}
\end{equation}
This is not surprising since, in presence of an excited vacuum, the QED coupling is modified as a result of the screening of the bare charge caused by the polarized cloud of virtual particles
\cite{steinhauser1998,sturm2013}. This effect can be calculated very precisely, up to 4-loop order for the leptonic contribution \cite{sturm2013}, and it has been verified experimentally as a function of the exchanged momentum \cite{kloe2017}. However, the calculation of the running of $\alpha$ with momentum does not include acceleration contributions. In the absence of a rigorous theory, we estimate this by approximating the vacuum polarization as $\Pi = \omega_p^2/\omega_0^2$, where $\omega_p = (e^2 n_p^{(v)}/\epsilon_0 m)^{1/2}$ is the plasma frequency of the virtual pairs, with $n_p^{(v)}$ the density of virtual particles and $\omega_0 \sim (k_\mathrm{B} T_{\rm U}/m)^{1/2}/\lambda_C$ the natural vibration frequency of the accelerated electron. The density of virtual pairs can be estimated using non-equilibrium QFT techniques based on the quantum Vlasov equation for fermions \cite{Kluger1998,Schmidt1998,Blaschke2009}; this gives
$n_p^{(v)} \sim m a^2 / 24 \pi^2 \hbar c^3$ \cite{Blaschke2006,Gregori2010}. The `running' of the fine structure constant with acceleration is then
\begin{equation}
    \alpha(a) = \frac{\alpha}{1+\Pi} \sim 
    \alpha \left(1- \frac{8 \pi^3 \alpha}{3} \frac{k_\mathrm{B} T_{\rm U}}{m c^2}\right).
\end{equation}
Up to a numerical constant, this is indeed the same scaling as in Eq.(\ref{alpha}).

\color{black}

Here we propose a variation of the experimental scheme suggested in Ref.~\cite{Crowley:2012t} as a possible way to directly measure the Unruh effect from accelerated electrons. Following directly our previous paper \cite{Crowley:2012t}, let us now consider the case of an electron oscillating in the field produced by a linearly polarized high-intensity laser beam. We can consider the acceleration to be constant if the laser frequency is such that (see Eq.~\ref{teq}):
\begin{equation}
\nu_\text{L} \lesssim \left( \frac{3 m^2 c^3 a}{64 \pi^4 h^2} \right)^{1/3} \approx 4.9 \times 10^{17} \left( \frac{I_\text{L}}{10^{20}\,\,\rm W/cm^2} \right)^{1/6} \,\,\rm Hz,
\end{equation}
where $I_\text{L}$ is the laser intensity. The above relation is well satisfied for all current high-intensity optical lasers.
Once the electron is accelerated the second step is to probe its motion using another laser beam. If the probe laser vector's potential is denoted as $\bf A$, the effective interaction Hamiltonian (including the quantum recoil correction seen above) describing the Thomson scattering process is \cite{Crowley:2012t} 
\begin{equation}
{\cal H}_{\rm int}({\bf r}) = \frac{e^2}{2 m} \left(1 - \eta \frac{k_\mathrm{B} T_\text{U}}{m c^2} \right)^2 {\bf A} \cdot {\bf A} \sim \frac{e^2}{2 m} {\rm e}^{-\mathbfit{q} \cdot \mathbfit{r}} {\bf A} \cdot {\bf A},
\label{ham}
\end{equation}
where we have used ${\mathbfit{a}} \sim {\mathbfit{r}}/t_{acc}^2$ (with $\mathbfit{r}$ the position vector of the electron), such that 
${\mathbfit{q}}=2 \eta k_\mathrm{B} T_\text{U} \mathbfit{a}/m c^4$. 
\color{black} To ensure non-relativistic motion, the probe frequency must be high enough to sample time scales that are shorter than the acceleration time, that is $c/\lambda_p \gtrsim t_\mathrm{acc}^{-1}$ (where $\lambda_p$ is the probe laser wavelength). \color{black} In practice, this implies that the probe must be operated at X-ray wavelengths, and X-ray free electron lasers \cite{Emma2010} are thus ideally suited for this.

\begin{figure}
    \centering
    \includegraphics[width=0.8\linewidth]{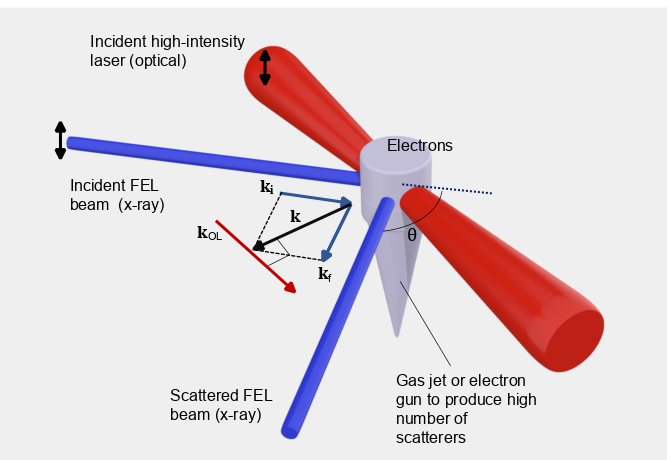}
    \caption{%
        Schematic of the setup for the proposed experiment to detect Unruh radiation. The black arrows indicate the polarization direction of the optical laser and FEL beams (both vertical in the figure), and ${\bf k}_{\rm OL}$ is the waveumber of the high-intensity optical laser.}
    \label{exp_fig2}
    \end{figure}

We now take an ensemble of uncorrelated electrons at a temperature $T$, all placed inside the high-power laser spot where they are accelerated as seen in Figure~\ref{exp_fig2} showing the schematic of the suggested experimental setup. Because of the modified Hamiltonian, the differential cross section for Thomson scattering which is the quantity which is measured in the experiment, is proportional to the dynamic structure factor, $S({\bf k}+i\mathbfit{q},\omega)$. This is given by \cite{Crowley:2012t,Gregori:2016apx}
\begin{equation}
S({\bf k}+i\mathbfit{q},\omega) = \int S_0({\bf k},\omega') F({\bf k}+i\mathbfit{q},\omega-\omega') d\omega',
\end{equation}
where ${\bf k}={\bf k_i}-{\bf k_f}$ is the scattering wavenumber, that is, the difference between the wavenumber of the incoming x-ray photons and the wavenumber of the scattered ones as shown in Figure \ref{exp_fig2}, and \cite{Crowley:2012t}
\begin{equation}
    F({\bf k}+i\mathbfit{q},\omega) = \sqrt{\frac{a}{\pi q}} \exp \left[ \frac{(\omega - {\bf k} \cdot {\bf v})^2}
    {q a}\right]. 
\end{equation}
Here, $\bf v$ is the bulk velocity of the electrons.
For a classical gas in equilibrium at a temperature $T$ we have \cite{Crowley:2012t},
\begin{eqnarray}
S_0({\bf k},\omega) = 
\left( \frac{m}{2 \pi k^2 k_\mathrm{B} T} \right)^{1/2} \exp \left[ -\frac{m}{2 k^2 k_\mathrm{B} T} \left(\omega - \frac{\hbar k^2}{2 m} \right)^2 \right].  
\end{eqnarray}

The electron motion induced by
the high-intensity optical laser remains along the acceleration direction, which also corresponds to the laser polarization axis. As shown in Figure~\ref{exp_fig2}, we can then choose the geometry of the scattering probe such that ${\bf k} \cdot {\bf v}=0$, and the total structure factor probed by the X-ray laser reduces to
\begin{eqnarray}
S({\bf k}+i\mathbfit{q},\omega) = 
\left( \frac{m}{2 \pi k^2 k_\mathrm{B} T_{\rm eff}} \right)^{1/2} \exp \left[ -\frac{m}{2 k^2 k_\mathrm{B} T_{\rm eff}} \left(\omega - \frac{\hbar k^2}{2 m} \right)^2 \right],  
\end{eqnarray}
implying that the modified Hamiltonian
(\ref{ham}) results in the electrons being probed by the Thomson scattering beam  exhibit an effective temperature given by \cite{Crowley:2012t}
\begin{equation}
    T_{\rm eff} = T\left( 1+ \frac{m a q}{2 k_\mathrm{B} k^2}\right) = 
    T\left[ 1+ 4 \eta \left( \frac{\pi k_\mathrm{B}}{\hbar k c} \right)^2 T_\text{U}^3 \right].
\end{equation}
Thus the probe measures a broadened electron velocity distribution
\begin{eqnarray}
   \frac{\Delta T}{T} = \frac{T_{\rm eff}-T}{T} \approx 
    0.5 \times \csc^2(\theta/2)
    \left( \frac{I_\text{L}}{10^{20}\,\,\rm W/cm^2} \right)^{3/2} \left(\frac{\lambda_p}{0.1\,\,\rm nm}\right)^2,
    \label{br}
\end{eqnarray}
where $\theta$ is the scattering angle. The expected broadening is within current experimental capabilities of detection \cite{Fletcher2015}.

We note that this uncertainty in the outgoing photon energy can also be interpreted as an uncertainty in the rest mass of the electron \cite{Gregori:2016apx}, that is,
by setting $ k_\mathrm{B} \Delta T = k_\mathrm{B} (T_{\rm eff} - T) = (\Delta m) c^2$, we obtain
\begin{equation}
    \Delta m = \frac{m a q}{2 k^2 c^2} = 
    4 \eta \left( \frac{\pi}{\hbar k c^2} \right)^2 (k_\mathrm{B} T_\text{U})^3.
    \label{dm}
\end{equation}
\noindent
We also note that, based on the discussions in the previous section,  taking $\hbar k c \sim k_\mathrm{B} T_\text{U}$, we can rewrite Eq.(\ref{dm}) as:
\begin{equation}
    \Delta m \sim \frac{4 \pi^2 \eta}{c^2} k_\mathrm{B} T_\text{U}.
\end{equation}

Thus 
if the measurement reveals a broadening as in Eq.(\ref{br}) it would provide direct evidence of the fact that the probe photons are scattering off electrons that are dressed by an excited vacuum.
\color{black}
The proposed experiment will thus provide a direct measurement of those quantum effects associated with the Unruh process that cannot be ascribed to just classical Larmor radiation.
We believe that the experimental approach suggest here satisfies all the requirements stated in Ref.~\cite{ROSU_1994} as needed for a robust claim of the quantum Unruh effect.

\color{black}

\section*{Conclusions}
The significance of experimentally detecting the Unruh effect can hardly be overstated. It will be a major scientific milestone marking a new era of experimental tests of quantum gravity as it will illuminate fundamental issues related to the (im)possibility of information loss \color{blue}\cite{Chen2017} \color{black} and the smallest possible scales in Nature. 

We note that such experiments at future high-intensity laser facilities may have impact on diverse technology applications such as quantum computing (as a quantum process under unscreenable gravitational fluctuations).


\section*{Acknowledgements}
GG would like to thank Dr Antonino di Piazza (University of Rochester) \color {black} Konstantin Beyer (MPIK) and Georgios Vacalis (University of Oxford) \color{black} for several discussions regarding this topic. CW is grateful to the Cruickshank Trust for support.


\bibliography{references}

\end{document}